\documentclass{PoS}

\title{AGILE observations of PSR B1509--58: a new class of highly magnetized,
'soft' gamma-ray pulsars?}

\ShortTitle{AGILE observations of PSR B1509--58}

\author{\speaker{Maura Pilia}\thanks{On behalf of the AGILE Team and AGILE
        Pulsar Working Group.}\\
        Dipartimento di Fisica, Universit\`a dell'Insubria, via Valleggio 11, I-22100, Como, Italy\\
        E-mail: \email{mpilia@oa-cagliari.inaf.it}}

\author{Alberto Pellizzoni\\
        INAF--Osservatorio Astronomico di Cagliari, localit\`a Poggio dei Pini, strada 54, I-09012 Capoterra, Italy \\
        E-mail: \email{apellizz@oa-cagliari.inaf.it}}

\def\agile {\emph{AGILE}}

\def\comp {\emph{COMPTEL}}

\def\lum {\mbox{erg s$^{-1}$}}

\def\psr {PSR~B1509--58}

\abstract{We present the results of $\sim$2.5 years \agile\ observations of
  \psr.  
The modulation significance of the light-curve above 30 MeV is at a 
5$\sigma$ confidence level and the light-curve is similar to those 
found earlier up to 30 MeV by \comp: a broad asymmetric first peak reaching 
its maximum $0.39 \pm 0.02$ cycles after the radio peak plus a second peak at
$0.94 \pm 0.03$.  
The gamma-ray spectral energy distribution of pulsed flux is well
described by a power-law (photon index $\alpha=1.87\pm0.09$) with a remarkable 
cutoff at $E_c=81\pm 20$~MeV, representing the softest
spectrum observed among $\gamma$-ray pulsars so far.
The unusual soft break in the spectrum of \psr\ has been interpreted 
in the framework of polar cap models as a signature of the exotic photon
splitting process in the strong magnetic field of this pulsar. 
In the case of an outer-gap scenario, or the two pole caustic model, better
constraints on the geometry of the emission would be needed from the radio band
in order to establish whether the conditions required by the models to
reproduce \agile\ light-curves and spectra match the polarization
measurements.} 

\FullConference{25th Texas Symposium on Relativistic Astrophysics - TEXAS 2010\\
		December 06-10, 2010\\
		Heidelberg, Germany}

\begin{document}

\section{Introduction}
\psr\ was discovered as an X-ray pulsar with the {\it Einstein} satellite 
and soon also detected at radio frequencies
(Manchester et al. 1982),  with  a  derived distance
supporting the  association with  the SNR MSH~15-52 ($d\sim 5.2$~kpc).  
With a period $P\simeq  150$~ms and a period derivative
$\dot P  \simeq 1.53 \times 10^{-12}$s~s$^{-1}$,  assuming the standard dipole
vacuum model, the estimated spin-down 
age for  this pulsar is 1570  years and  its inferred  surface
magnetic field is one of the  highest observed for an ordinary radio pulsar:
$B=3.1\times 10^{13}$~G, as calculated at the pole. Its rotational energy  loss
rate is $\dot E = 1.8 \times  10^{37}$~erg/s. 

The young age  and the  high rotational  energy loss rate made this
pulsar a promising target for the gamma-ray
satellites. In fact, the instruments on board of the
{\it Compton Gamma-Ray Observatory} ({\it CGRO}) observed its
pulsation at low gamma-ray energies,
but it was not detected with high
significance by the {\it Energetic Gamma-Ray Experiment Telescope} ({\it
  EGRET}), the instrument  operating at  the energies from 30~MeV to 30~GeV.
This was remarkable, since all other known gamma-ray pulsars show spectral
turnovers well above 100~MeV (e.g. Thompson 2004).  
Harding et al. (1997) suggested that the break in the spectrum  
could be interpreted as due to inhibition of the pair-production
caused by the photon-splitting phenomenon (Adler et al. 1970). The photon
splitting appears, in the frame of the polar cap models, in
relation with a very high magnetic field. 
An alternative explanation is proposed by Zhang \& Cheng (2000) using a
three dimensional outer gap model. They propose that the gamma-ray emission is
produced by synchrotron-self Compton radiation above the outer gap.

The Italian satellite \agile\ (Tavani et al. 2009) obtained the first
  detection  of  \psr\ in the {\it EGRET} band (Pellizzoni et al. 2009b)
  confirming the occurrence of a spectral break. 
Here we summarize the results of a $\sim 2.5$~yr monitoring campaign
of \psr\ with \agile, improving counts statistics, and therefore
ligh-tcurve characterization, with respect to earlier \agile\ observations.
More details on this analysis can be found in Pilia et al. (2010). 
With these observations the spectral energy distribution (SED) at
$E<300$~MeV, where the remarkable spectral turnover is observed, can be
  assessed. 

\section{\emph{AGILE} Observations, Data Analysis and Results}

\agile\ devoted a large amount of observing time to the region of \psr.
For details on \agile\ observing strategy, timing calibration and gamma-ray
pulsars analysis the reader can refer to Pellizzoni et al. (2009a,b).
A total exposure of $3.8
\times 10^{9}$~cm$^2$~s ($E > 100$ MeV) was obtained during the $2.5$~yr
period of observations (July 2007 - October 2009) which,
combined with \agile\ effective area, gives our observations a good
photon harvest from this pulsar.

Simultaneous radio observations of \psr\ with the
Parkes radiotelescope in Australia are ongoing since the epoch of
\agile's launch.
Strong timing noise was present
and it was accounted for using the
$fitwaves$ technique developed in the framework of the TEMPO2 radio
timing software (Hobbs et al. 2004, 2006).   
Using the radio ephemeris provided by the {\it Parkes} telescope, 
we performed the folding of the gamma-ray light-curve including the wave
terms (Pellizzoni et al. 2009a).  
An optimized analysis followed, 
aimed at cross-checking and maximization 
of the significance of the detection, including an energy-dependent events 
extraction angle around the source position based on the instrument
point-spread-function (PSF). 
The chi-squared ($\chi^2$)-test applied to the 10 
bin light-curve at $E>30$ MeV  gave a detection significance of $\sigma = 4.8$.
The unbinned $Z_n^2$-test gave a significance of $\sigma = 5.0$ with $n=2$
harmonics.
The difference between the radio and gamma-ray ephemerides was 
$\Delta P_{radio,\gamma}=10^{-9}$~s, at a level lower than the error in
the parameter, showing perfect agreement
among radio and gamma-ray ephemerides as
expected, further supporting our detection and \agile\ timing calibration.  

\begin{figure}
\centering
\includegraphics[width=7cm]{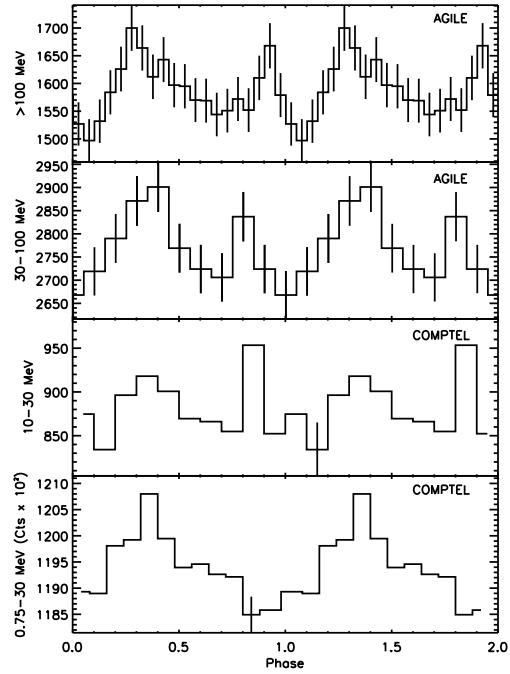} 
\caption{\label{fig:lc_tot}
Phase-aligned gamma-ray light-curves of \psr\ with radio peak at
phase 0.
From the top: \agile\
 $>100$~MeV, 20 bins, 7.5
ms resolution; \agile\ $<100$~MeV, 10 bins, 15 ms
resolution; 
\comp\ 10--30~MeV and
\comp\ 0.75--30~MeV (from Kuiper et al. (1999).} 

\end{figure}

We observed \psr\ in three energy bands: 30--100~MeV, 100--500~MeV and above
500~MeV. 
We did not detect pulsed emission at a significance $\sigma \geq 2$ for $E >
500$~MeV. 
The $\gamma$-ray light-curves of \psr\ for different energy bands
are shown in Fig. \ref{fig:lc_tot}. 
The \agile\ $E>30$ MeV light-curve shows two peaks at phases
$\phi_1  = 0.39 \pm 0.02$ and $\phi_2  = 0.94 \pm 0.03 $ with respect to the
single radio peak, here put at phase 0.
The phases are calculated using a Gaussian fit to the peaks, yielding
a FWHM of $0.29(6)$ for the first peak and of $0.13(7)$ for the second peak,
where we quote in parentheses (here and throughout the paper) the 1$\sigma$ error
on the last digit.
The first peak is coincident in phase with \comp's peak (Kuiper et al. 1999). In its
highest energy band (10--30~MeV) \comp\ showed the indication of a second peak 
(even though the modulation had low significance, $2.1 \sigma$).
This second
peak is coincident in phase with \agile's second peak (Fig. \ref{fig:lc_tot}).
\agile\ thus confirms the previously marginal detection of a second peak.

Based on our exposure 
we derived the $\gamma$-ray flux from the number of
pulsed counts. 
The pulsed fluxes in the three \agile\ energy bands were 
$F_{\gamma}= 10(4)\times 10^{-7}
$~ph~cm$^{-2}$~s$^{-1}$ in the 30--100~MeV band, 
$F_{\gamma}= 2.1(5)\times 10^{-7} $~ph~cm$^{-2}$~s$^{-1}$ in the 100--500~MeV
band  
and a $1 \sigma$ upper limit $F_{\gamma}< 8\times 10^{-8}
$~ph~cm$^{-2}$~s$^{-1}$ for $E>500$~MeV.

\begin{figure}
\centering
\includegraphics[width=10cm]{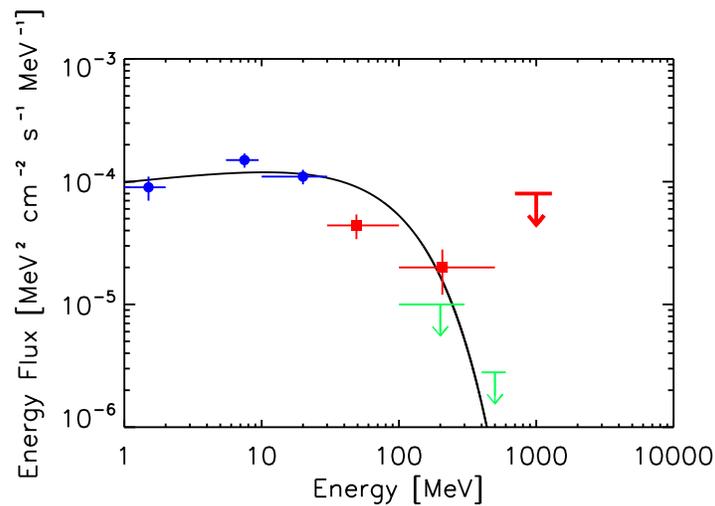}
\caption{\label{spec} SED of \psr\ (solid
 line) obtained from a fit of pulsed fluxes from soft to hard
  gamma-rays. The three round
   points represent \comp\ observations (Kuiper et al. 1999). The two square points
  represent 
  \agile\ pulsed flux in two bands ($30<E<100$~MeV and $100<E<500$~MeV). 
The thicker arrow represents
  \agile\ upper limit above 500~MeV.  The two thin arrows represent {\it
  Fermi} upper limits from Adbo et al. (2010b)
  }
\end{figure}

Fig. 2 shows the SED of \psr\ based on
\agile's and \comp's observed fluxes. {\it Fermi} upper limits are also shown,
which are consistent with our measurements at a $2\sigma$ confidence level.
\comp\ observed this pulsar in three energy bands: 0.75--3~MeV,
3--10~MeV, 10--30~MeV, suggesting a 
spectral break between 10 and 30 MeV. 
\agile\ pulsed flux 
confirms the presence of a soft spectral break.
As shown in Fig. \ref{spec}, we modeled
the observed \comp\ and \agile\ fluxes with a power-law plus cutoff fit 
using the Minuit minimization package (James et al. 1975): $F(E)=k \times
E^{-\alpha}\exp[-(E/E_{c})^{\beta}]$, 
with three free parameters: the normalization $k$, the spectral index
$\alpha$, the cutoff energy $E_c$ and allowing $\beta$ to assume values of 1
and 2 (indicating either an exponential or a superexponential cutoff). 
No acceptable $\chi^2$ values were obtained for a superexponential cutoff, the
presence of which can be excluded at a
$3.5\sigma$ confidence level,
while for an exponential cutoff we found $\chi^2_{\nu}=3.2$ for $\nu = 2$
degrees of freedom, corresponding to a
null hypothesis probability of 0.05. 
The best values thus obtained for the parameters of the fit were:
$k=1.0(2)\times 10^{-4}$~s$^{-1}$~cm$^{-2}$, $\alpha=1.87(9)$, $E_{c}=81(20)$~MeV.

\section{Discussion}

The bulk of the spin-powered pulsar flux is usually emitted in the MeV-GeV
energy band with  
spectral breaks at $\leq 10$~GeV (e.g. Abdo et al. 2010a).
\psr\ has the softest spectrum observed among gamma-ray
pulsars, with a sub-GeV cutoff at $E \approx 80$~MeV. 
In the following we discuss how the new \agile\ observations can constrain the
models for emission from the pulsar magnetosphere (for an extended discussion see Pilia et al. 2010).

When \psr\ was detected in soft gamma-rays but not significantly at $E>30$~MeV,
it was proposed that the mechanism 
responsible for this low-energy spectral break might be photon splitting
(Harding et al. 1997).
The photon splitting (Adler et al. 1970) is an exotic third-order quantum
electro-dynamics 
process expected when the 
magnetic field approaches or exceeds the $critical$ value defined as
$B_{cr}=m^2_e c^3/(e\hbar)=4.413\times 10^{13}$~G. 
Most current theories for
the generation of coherent radio 
emission in pulsar magnetospheres require formation of an
electron-positron pair plasma developing via electromagnetic cascades. In very
high magnetic fields the formation of pair cascades can be altered
by the process of photon splitting: $\gamma \rightarrow \gamma\gamma$, 
which will operate as an 
attenuation mechanism in the high-field regions near pulsar polar caps. 
Since it has no energy threshold, photon splitting can attenuate photons below
the threshold for pair production, 
thus determining a spectral cutoff at lower energies.

In the case of \psr\ a polar cap model with photon splitting would be
able to explain the soft gamma-ray emission and the low energy
spectral cutoff, now quantified by \agile\ observations.
Based on the observed cutoffs, which are related to the photons' saturation
escape energy, 
we can derive constraints on the magnetic field strength at emission,
in the framework of photon splitting:
\begin{equation}
\epsilon_{esc}^{sat} \simeq 0.077(B^{\prime}  \sin \theta_{kB,0})^{-6/5} 
\label{eq:emax}
\end{equation}
where $\epsilon_{esc}^{sat}$ is the photon saturation escape energy, $B^{\prime}=B/B_{cr}$ and
$\theta_{kB,0} $ is the angle between the 
photon momentum and the magnetic field vectors at the surface and is here
assumed to be very small: 
 $\theta_{kB,0} \lesssim 0.57 ^{\circ} $
(see Harding et al. 1997). 
Using the observed energy cutoff ($\epsilon_{esc}^{sat} \simeq E= 80$~MeV) we
find that $B^{\prime} \gtrsim 0.3$, which 
implies an emission altitude $\lesssim 1.3 R_{NS}$, which is the height where
possibly also pair production could ensue. 
This altitude of emission agrees with the polar cap models
(see e.g. Daugherty \& Harding 1996). A smaller energy cutoff, as in
Harding et al. (1997), would have implied even lower emission altitude and a
sharper break, possibly caused by the total absence of pair production. 
It is apparent that small differences in the emission position will cause
strong differences in  spectral shape. This is possibly the reason for 
the different emission properties of the two peaks as observed in the total
(\agile\ plus \comp)  
gamma-ray energy band. Also, a trend can be observed, from lower to higher
energies (see the X-ray light-curve for the trend in the first peak, as in
Fig. 3 of Kuiper et al. 1999), of the peaks slightly drifting away from the
radio peak. This we assume 
to be another signature of the fact that small variations in emission height
can be responsible for sensible changes in the light-curves
in such a high magnetic field.
The scenario proposed by Harding et al. (1997) 
is strengthened by its prediction that PSR~B0656+14
should have a cutoff with an intermediate value between \psr\ and the
other gamma-ray pulsars. 
Additionally, \psr\ (Kuiper et al. 1999, Crawford et al. 2001) and PSR~B0656+14 
(De Luca et al. 2005, Weltevrede et al. 2010) show
evidence of an aligned geometry, which could imply polar cap emission. 

The polar cap model 
as an emission mechanism is debated.
From the theoretical point of view, the angular momentum is
not conserved in polar cap emission (Cohen \& Treves 1972,
Holloway 1977, Treves et al. 2010).
And a preferential explanation of the observed gamma-ray
light-curves with high altitude cascades 
comes from the recent
results by {\it Fermi} (Abdo et al. 2010a).
In the case of \psr, 
the derived gamma-ray luminosity from the flux at 
$E>1$~MeV, considering a 1~sr beam sweep is $L_{\gamma}=4.2^{+0.5}_{-0.2}
d^2_{5.2} \times 
10^{35}$~erg/s, where $d_{5.2}$ indicates the distance in units of 5.2 
kpc.
While traditionally the beaming fraction ($f_{\Omega}$) was considered to be
the equivalent of a 1~sr sweep, nowadays (see e.g. Watters et al. 2009)
the tendency is to consider a larger beaming
fraction ($f_{\Omega} \approx 1$), close to a $4\pi$~sr beam. 
Using $f_{\Omega}=1$ in our calculations,
we would have obtained $L_{\gamma}=5.8^{+0.1}_{-0.8}d^2_{5.2}\times
10^{36}$~\lum. 
Thus the maximum conversion efficiency of the rotational
energy loss ($\dot E \approx 1.8 \times  10^{37}$~\lum, see Section~1) 
into gamma-ray luminosity is 0.3.
Our result is not easily comparable with the typical gamma-ray luminosities
above 100~MeV, 
because for \psr\ this energy band is beyond the spectral break.
Using \agile\ data alone we obtained a luminosity above 30~MeV
$L_{\gamma}=5.2(6) d^2_{5.2}\times  10^{35}$~erg/s, again for a 1~sr beam.
If the gamma-ray 
luminosity cannot account for a large fraction of the rotational energy loss,
then the 
angular momentum conservation objection from Cohen \& Treves (1972) becomes less
cogent for this pulsar, 
exactly as it happens for the radio emission. 

Alternatively, if such an efficiency as that of \psr\ were incompatible with
this conservation law, an interpretation of \psr\ emission can be sought
 in the frame of the three dimensional outer gap model
 (Zhang \& Cheng 2000). 
According to their model, hard X-rays and low energy gamma-rays are both
produced by synchrotron self-Compton radiation of secondary 
e$^+$e$^-$ pairs of the outer gap. Therefore, as observed, the phase
offset of hard X-rays and low energy gamma-rays with respect to the radio
pulse is the same, with the possibility of a small lag due to the thickness of
the emission region. 
According to their estimates 
a magnetic inclination angle $\alpha\approx 60 ^o$ and a viewing
angle $\zeta \approx 75 ^o$ are
required to reproduce the observed light-curve. 
Finally, using the simulations of Watters et al. 2009), 
who produced a map of pulse profiles for different combinations of
angles $\alpha$ and $\zeta$ in the different emission models,
the observed light-curve from \agile\ is best reproduced
if $\alpha\approx 35 ^{\circ}$  and   $\zeta \approx 90
^{\circ}$, in the framework of the two pole caustic model
(Dyks \& Rudak 2003).

The values of $\alpha$ and $\zeta$ required by the Zhang \& Cheng model
are not in good 
agreement with the corresponding values obtained with radio measurements.
In fact, Crawford et al. (2001) observe that $\alpha$ must be $< 60 ^{\circ}$ 
at the $3 \sigma$ level.
The prediction obtained by the simulations of Watters et al. (2009)
better agrees with the radio polarization 
observations.
In fact, in the framework of the rotating vector model (RVM, see
e.g. Lorimer \& Kramer 2004 and references therein),
Crawford et al. (2001)  also propose that, 
if the restriction is imposed that $\zeta > 70 ^{\circ}$ (Melatos 1997),
then $\alpha > 30 ^{\circ}$ at the $3 \sigma$ level.
For these values, however, the Melatos model for the spin down of an oblique
rotator 
predicts a braking index $n>2.86$, slightly inconsistent with the observed 
value ($n=2.839(3)$, see Livingstone et al. 2005).
Also in the case of
PSR~B0656+14, Weltevrede et al. (2010) conclude that 
the large values of
$\alpha$ and $\zeta$ are somewhat at odds with the constraints from the
modeling of the radio data and the thermal X-rays which seem to imply a more
aligned geometry. 
Improved radio polarization measurements would help placing better constraints
on the pulsar geometry and therefore on the possibility of a gap in the
extended or outer magnetosphere, but the quality of the polarization
measurements from Crawford et al. (2001) is already excellent, 
the problem being that \psr, like most pulsars, only shows
emission over a limited pule phase range and therefore the RVM models
are highly degenerate.

At present the geometry privileged by the
state of the art measurements is best compatible with polar cap models. 
Higher statistics in the number of observed gamma-ray pulsars  could help
characterize a class of "outliers" having gamma-ray emission from the
polar caps, which potentially constitute a privileged target for \agile.

\end{document}